\documentclass{sig-alternate}

\usepackage[
pdftitle={Complex Adaptive Digital EcoSystems},
pdfauthor={Gerard Briscoe},
pdfsubject={Digital Ecosystems},
pdfkeywords={Digital Ecosystems},
colorlinks=true, linkcolor=black, citecolor=black, filecolor=black, urlcolor=black, hyperfootnotes=false]{hyperref} 

\usepackage{graphicx, acronym, ifthen, substr, color}

\usepackage[paper=a4paper, asymmetric, left=2cm,right=1.4cm,top=2cm,bottom=4.2cm]{geometry}

\newboolean{final}\setboolean{final}{true}

\newboolean{arxiv}\setboolean{arxiv}{false}

\ifthenelse{\boolean{final}}{}{\usepackage{showlabels}\usepackage{citeext}}

\ifthenelse{\boolean{arxiv}}
	{\newcommand{\arxivfootnote}{\footnote}}
	{\def\arxivfootnote#1{}}

\ifthenelse{\boolean{arxiv}}
	{\def\arxivOnly#1{#1}}
	{\def\arxivOnly#1{}}

\ifthenelse{\boolean{arxiv}}
	{\def\arxivNot#1{}}
	{\def\arxivNot#1{#1}}

\ifthenelse{\boolean{arxiv}}
	{}
	{}

\newcommand{\tfigure}[9]
 {
 \IfSubStringInString{!}{#7}{\begin{figure}[#7]}{\begin{figure}[!t]}
 \IfSubStringInString{mm}{#8}{\vspace{#8}}{}
 \centering
 
 \includegraphics[#1]{images/#2.pdf}
 
 \vspace{#6}
 \caption[#4]
 {
 \label{#2}
 #4: #5
 }
 \IfSubStringInString{mm}{#9}{\vspace{#9}}{}
 \end{figure}
 }

\definecolor{tred}{RGB}{255,0,0}

\acrodef{DEAL}{Digital Ecosystem for Agriculture \& Rural Livelihood}
\acrodef{PCG}{Projected Conjugate Gradient} 
\acrodef{QP}{quadratic programming}
\acrodef{RBF}{Radial-Basis Function}
\acrodef{ABM}{Agent-Based Modelling}
\acrodef{AI}{Artificial Intelligence}
\acrodef{DAI}{Distributed Artificial Intelligence}
\acrodef{API}{Application Programming Interface}
\acrodef{ARF}{p14ARF human tumor-suppressor gene}
\acrodef{B2B}{business-to-business}
\acrodef{BDP}{Biological Design Pattern}
\acrodef{BGS}{Best Guess Solution}
\acrodef{BIC}{Biologically-Inspired Computing}
\acrodef{BML}{Business Modelling Language}
\acrodef{BPEL}{Business Process Execution Language}
\acrodef{BPMN}{Business Process Modelling Notation}
\acrodef{CAS}{Complex Adaptive Systems}
\acrodef{COBOL}{COmmon Business-Oriented Language}
\acrodef{DBE}{Digital Business Ecosystem}
\acrodef{DKE}{Digital Knowledge Ecosystem}
\acrodef{DE}{Digital Ecosystem}
\acrodef{DEC}{distributed evolutionary computing}
\acrodef{DGA}{Distributed genetic algorithms}
\acrodef{DIS}{Distributed Intelligence System}
\acrodef{DNA}{Deoxyribose Nucleic Acid}
\acrodef{DOP}{DBE Open Protocol}
\acrodef{DSS}{Distributed Storage System}
\acrodef{EAP}{Evolving Agent Population}
\acrodef{ebXML}{e-business eXtensible Markup Language}
\acrodef{EC}{Evolutionary Computing}
\acrodef{ECJ}{Evolutionary Computing in Java}
\acrodef{EE}{Evolutionary Environment}
\acrodef{EFL}{Evolutionary Framework for Language}
\acrodef{FLE}{Framework for Language Ecosystems}
\acrodef{EOA}{Ecosystem-Oriented Architecture}
\acrodef{ESS}{evolutionary stable strategy}
\acrodef{EvE}{Evolutionary Environment}
\acrodef{ExE}{Execution Environment}
\acrodef{FCB}{Framework for Computational Biomimicry}
\acrodef{FFF}{Fitness Function Framework}
\acrodef{FL}{Fitness Landscape}
\acrodef{HWU}{Heriot-Watt University}
\acrodef{ICL}{Imperial College London}
\acrodef{ICT}{Information and Communications Technology}
\acrodef{INTEL}{Intel Ireland}
\acrodef{IPA}{International Phonetic Alphabet}
\acrodef{ISUFI}{Istituto Superiore Universitario di Formazione Interdisciplinare}
\acrodef{JDJ}{Java Developer's Journal}
\acrodef{KC}{Kolmogorov-Chaitin}
\acrodef{LAN}{local area network}
\acrodef{LSE}{London School of Economics and Political Science}
\acrodef{MAS}{Multi-Agent System}
\acrodef{MDL}{Minimum Description Length}
\acrodef{MDM2}{murine double minute 2}
\acrodef{MFT}{Mean Field Theory}
\acrodef{MoAS}{Mobile Agent System}
\acrodef{MOF}{Meta Object Facility}
\acrodef{MUH}{migration and usage history}
\acrodef{NIC}{Nature Inspired Computing}
\acrodef{NN}{Neural Network}
\acrodef{NoE}{Network of Excellence}
\acrodef{OMG}{Open Mac Grid}
\acrodef{OPAALS}{Open Philosophies for Associative Autopoietic Digital Ecosystems}
\acrodef{P2P}{peer-to-peer}
\acrodef{P53}{protein 53}
\acrodef{PDA}{Personal Digital Assistant}
\acrodef{QoS}{quality of service}
\acrodef{REST}{REpresentational State Transfer}
\acrodef{RNA}{Deoxyribose Nucleic Acid}
\acrodef{SAE}{Software Agent Ecosystem}
\acrodef{SBML}{Systems Biology Modelling Language}
\acrodef{SBVR}{Semantics of Business Vocabulary and Business Rules}
\acrodef{SDL}{Service Description Language}
\acrodef{SF}{Service Factory}
\acrodef{SIM}{Social Interaction Mechanism}
\acrodef{SM}{Service Manifest}
\acrodef{SME}{Small and Medium sized Enterprise}
\acrodef{SML}{Service Modelling Language}
\acrodef{SMO}{Sequential Minimal Optimisation}
\acrodef{SOA}{Service-Oriented Architecture}
\acrodef{SOAP}{Simple Object Access Protocol}
\acrodef{SOC}{Self-Organised Criticality}
\acrodef{SOLUTA}{SOLUTA.NET}
\acrodef{SOM}{Self-Organising Map}
\acrodef{SSL}{Semantic Service Language}
\acrodef{STU}{Salzburg Technical University}
\acrodef{SUN}{Sun Microsystems}
\acrodef{SVM}{Support Vector Machine}
\acrodef{TM}{Turing Machine}
\acrodef{UBHAM}{University of Birmingham}
\acrodef{UDDI}{Universal Description Discovery and Integration}
\acrodef{UML}{Unified Modelling Language}
\acrodef{URI}{Uniform Resource Identifier}
\acrodef{UTM}{Universal Turing Machine}
\acrodef{VLP}{variable length population}
\acrodef{VLS}{variable length sequences}
\acrodef{vls}{variable length sequence}
\acrodef{WP}{Work-Package}
\acrodef{WSDL}{Web Services Definition Language}
\acrodef{XMI}{XML Metadata Interchange}
\acrodef{XML}{eXtensible Markup Language}
\acrodef{MD5}{Message-Digest algorithm 5}
\acrodef{GA}{genetic algorithm}
\acrodef{GP}{genetic programming}
\acrodef{MASON}{Multi-Agent Simulator Of Neighbourhoods}
\acrodef{Repast}{Recursive Porous Agent Simulation Toolkit}
\acrodef{JCLEC}{Java Computing Library for Evolutionary Computing}
\acrodef{OWL-S}{Web Ontology Language - Service}
\acrodef{EGT}{Evolutionary Game Theory}
\acrodef{RBF}{Radial Basis Functions}
\acrodef{SWS}{Semantic Web Services}
\acrodef{HDD}{Hard Disk Drive}
\acrodef{SSD}{Solid-State Drive}

\acrodef{im1}{are different probabilities of going from island 1 to island 2, as there is of going from island 2 to island 1.}
\acrodef{im2}{mirrors the naturally inspired quality that although two populations have the same physical separation, it may be easier to migrate in one direction than the other, i.e. fish migration is easier downstream than upstream.}
\acrodef{digEco}{with the agents, the populations, the agent migration for \acl{DEC}, and the environmental selection pressures provided by the user base, then the union of the habitats creates the Digital Ecosystem}
\acrodef{archComTop}{many strongly connected clusters (communities), called {sub-networks} (quasi-complete graphs), with a few connections between these clusters (communities). Graphs with this topology have a very high clustering coefficient and small characteristic path lengths}
\acrodef{similarCap}{requests are evaluated on separate {islands} (populations), and so adaptation is accelerated by the sharing of solutions between evolving populations (islands), because they are working to solve similar requests (problems).}
\acrodef{picUser}{will formulate queries to the Digital Ecosystem by creating a request as a {semantic description}, like those being used and developed in \acp{SOA}}
\acrodef{picUserReq}{A population is then instantiated in the user's habitat in response to the user's request, seeded from the agents available at their habitat.}
\acrodef{urlCapUnifrom}{The {observed} frequencies of the application (agent aggregation) size mostly matched the {expected} frequencies}
\acrodef{urlCapGaussian}{The {observed} frequencies of the application (agent aggregation) size matched the {expected} frequencies with only minor variations}
\acrodef{urlpower}{The {observed} frequencies of the application (agent aggregation) size matched the {expected} frequencies with some variation}
\acrodef{urvunifromCap}{The {observed} frequencies for the number of agent attributes mostly matched the {expected} frequencies}
\acrodef{urvgaussianCap}{The {observed} frequencies for the number of agent attributes again followed the {expected} frequencies, but there was variation}
\acrodef{urvpowerCap}{The {observed} frequencies for the number of agent attributes also followed the {expected} frequencies, but there was variation}
\acrodef{im1}{are different probabilities of going from island 1 to island 2, as there is of going from island 2 to island 1.}
\acrodef{im2}{mirrors the naturally inspired quality that although two populations have the same physical separation, it may be easier to migrate in one direction than the other, i.e. fish migration is easier downstream than upstream.}
\acrodef{digEco}{with the agents, the populations, the agent migration for \acl{DEC}, and the environmental selection pressures provided by the user base, then the union of the habitats creates the Digital Ecosystem}
\acrodef{archComTop}{many strongly connected clusters (communities), called {sub-networks} (quasi-complete graphs), with a few connections between these clusters (communities). Graphs with this topology have a very high clustering coefficient and small characteristic path lengths}
\acrodef{similarCap}{requests are evaluated on separate {islands} (populations), and so adaptation is accelerated by the sharing of solutions between evolving populations (islands), because they are working to solve similar requests (problems).}
\acrodef{picUser}{will formulate queries to the Digital Ecosystem by creating a request as a {semantic description}, like those being used and developed in \acp{SOA}}
\acrodef{picUserReq}{A population is then instantiated in the user's habitat in response to the user's request, seeded from the agents available at their habitat.}
\acrodef{urlCapUnifrom}{The {observed} frequencies of the application (agent aggregation) size mostly matched the {expected} frequencies}
\acrodef{urlCapGaussian}{The {observed} frequencies of the application (agent aggregation) size matched the {expected} frequencies with only minor variations}
\acrodef{urlpower}{The {observed} frequencies of the application (agent aggregation) size matched the {expected} frequencies with some variation}
\acrodef{urvunifromCap}{The {observed} frequencies for the number of agent attributes mostly matched the {expected} frequencies}
\acrodef{urvgaussianCap}{The {observed} frequencies for the number of agent attributes again followed the {expected} frequencies, but there was variation}
\acrodef{urvpowerCap}{The {observed} frequencies for the number of agent attributes also followed the {expected} frequencies, but there was variation}
\acrodef{im1}{are different probabilities of going from island 1 to island 2, as there is of going from island 2 to island 1.}
\acrodef{im2}{mirrors the naturally inspired quality that although two populations have the same physical separation, it may be easier to migrate in one direction than the other, i.e. fish migration is easier downstream than upstream.}
\acrodef{digEco}{with the agents, the populations, the agent migration for \acl{DEC}, and the environmental selection pressures provided by the user base, then the union of the habitats creates the Digital Ecosystem}
\acrodef{archComTop}{many strongly connected clusters (communities), called {sub-networks} (quasi-complete graphs), with a few connections between these clusters (communities). Graphs with this topology have a very high clustering coefficient and small characteristic path lengths}
\acrodef{similarCap}{requests are evaluated on separate {islands} (populations), and so adaptation is accelerated by the sharing of solutions between evolving populations (islands), because they are working to solve similar requests (problems).}
\acrodef{picUser}{will formulate queries to the Digital Ecosystem by creating a request as a {semantic description}, like those being used and developed in \acp{SOA}}
\acrodef{picUserReq}{A population is then instantiated in the user's habitat in response to the user's request, seeded from the agents available at their habitat.}
\acrodef{urlCapUnifrom}{The {observed} frequencies of the application (agent aggregation) size mostly matched the {expected} frequencies}
\acrodef{urlCapGaussian}{The {observed} frequencies of the application (agent aggregation) size matched the {expected} frequencies with only minor variations}
\acrodef{urlpower}{The {observed} frequencies of the application (agent aggregation) size matched the {expected} frequencies with some variation}
\acrodef{urvunifromCap}{The {observed} frequencies for the number of agent attributes mostly matched the {expected} frequencies}
\acrodef{urvgaussianCap}{The {observed} frequencies for the number of agent attributes again followed the {expected} frequencies, but there was variation}
\acrodef{urvpowerCap}{The {observed} frequencies for the number of agent attributes also followed the {expected} frequencies, but there was variation}
\acrodef{im1}{are different probabilities of going from island 1 to island 2, as there is of going from island 2 to island 1.}
\acrodef{im2}{mirrors the naturally inspired quality that although two populations have the same physical separation, it may be easier to migrate in one direction than the other, i.e. fish migration is easier downstream than upstream.}
\acrodef{digEco}{with the agents, the populations, the agent migration for \acl{DEC}, and the environmental selection pressures provided by the user base, then the union of the habitats creates the Digital Ecosystem}
\acrodef{archComTop}{many strongly connected clusters (communities), called {sub-networks} (quasi-complete graphs), with a few connections between these clusters (communities). Graphs with this topology have a very high clustering coefficient and small characteristic path lengths}
\acrodef{similarCap}{requests are evaluated on separate {islands} (populations), and so adaptation is accelerated by the sharing of solutions between evolving populations (islands), because they are working to solve similar requests (problems).}
\acrodef{picUser}{will formulate queries to the Digital Ecosystem by creating a request as a {semantic description}, like those being used and developed in \acp{SOA}}
\acrodef{picUserReq}{A population is then instantiated in the user's habitat in response to the user's request, seeded from the agents available at their habitat.}
\acrodef{urlCapUnifrom}{The {observed} frequencies of the application (agent aggregation) size mostly matched the {expected} frequencies}
\acrodef{urlCapGaussian}{The {observed} frequencies of the application (agent aggregation) size matched the {expected} frequencies with only minor variations}
\acrodef{urlpower}{The {observed} frequencies of the application (agent aggregation) size matched the {expected} frequencies with some variation}
\acrodef{urvunifromCap}{The {observed} frequencies for the number of agent attributes mostly matched the {expected} frequencies}
\acrodef{urvgaussianCap}{The {observed} frequencies for the number of agent attributes again followed the {expected} frequencies, but there was variation}
\acrodef{urvpowerCap}{The {observed} frequencies for the number of agent attributes also followed the {expected} frequencies, but there was variation}
\acrodef{im1}{are different probabilities of going from island 1 to island 2, as there is of going from island 2 to island 1.}
\acrodef{im2}{mirrors the naturally inspired quality that although two populations have the same physical separation, it may be easier to migrate in one direction than the other, i.e. fish migration is easier downstream than upstream.}
\acrodef{digEco}{with the agents, the populations, the agent migration for \acl{DEC}, and the environmental selection pressures provided by the user base, then the union of the habitats creates the Digital Ecosystem}
\acrodef{archComTop}{many strongly connected clusters (communities), called {sub-networks} (quasi-complete graphs), with a few connections between these clusters (communities). Graphs with this topology have a very high clustering coefficient and small characteristic path lengths}
\acrodef{similarCap}{requests are evaluated on separate {islands} (populations), and so adaptation is accelerated by the sharing of solutions between evolving populations (islands), because they are working to solve similar requests (problems).}
\acrodef{picUser}{will formulate queries to the Digital Ecosystem by creating a request as a {semantic description}, like those being used and developed in \acp{SOA}}
\acrodef{picUserReq}{A population is then instantiated in the user's habitat in response to the user's request, seeded from the agents available at their habitat.}
\acrodef{urlCapUnifrom}{The {observed} frequencies of the application (agent aggregation) size mostly matched the {expected} frequencies}
\acrodef{urlCapGaussian}{The {observed} frequencies of the application (agent aggregation) size matched the {expected} frequencies with only minor variations}
\acrodef{urlpower}{The {observed} frequencies of the application (agent aggregation) size matched the {expected} frequencies with some variation}
\acrodef{urvunifromCap}{The {observed} frequencies for the number of agent attributes mostly matched the {expected} frequencies}
\acrodef{urvgaussianCap}{The {observed} frequencies for the number of agent attributes again followed the {expected} frequencies, but there was variation}
\acrodef{urvpowerCap}{The {observed} frequencies for the number of agent attributes also followed the {expected} frequencies, but there was variation}

\usepackage{color}
\definecolor{gray}{rgb}{0.5,0.5,0.5}

\toappear{}

\begin{document}
\conferenceinfo{MEDES'10}{October 26-29, 2010, Bangkok, Thailand}
\CopyrightYear{2010} 
\crdata{978-1-4503-0047-6/10/10}  

\title{Complex Adaptive Digital EcoSystems}

\numberofauthors{1} 
\author{
\alignauthor
Gerard Briscoe\\
 \affaddr{Intelligent Systems Lab}\\
 \affaddr{Department of Computer Science}\\
 \affaddr{Heriot Watt University}\\
 \affaddr{United Kingdom}\\
 \email{g.briscoe@hw.ac.uk}
}

\maketitle
\begin{abstract}

We investigate an abstract conceptualisation of Digital Ecosystems from a computer science perspective. We then provide a conceptual framework for the cross pollination of ideas, concepts and understanding between different classes of ecosystems through the universally applicable principles of \ac{CAS} modelling. A framework to assist the cross-disciplinary collaboration of research into Digital Ecosystems, including \acp{DBE} and \acp{DKE}. So, we have defined the key steps towards a theoretical framework for Digital Ecosystems, that is compatible with the diverse theoretical views prevalent. Therefore, a theoretical edifice that can unify the diverse efforts within Digital Ecosystems research.

\end{abstract}


\section{Introduction}
Conceptualising ecosystems has been an inherent part of efforts, which presents us with an opportunity to formalise our current and future efforts to improve the cross-disciplinary knowledge transfer required. 

\tfigure{width=3.33in}{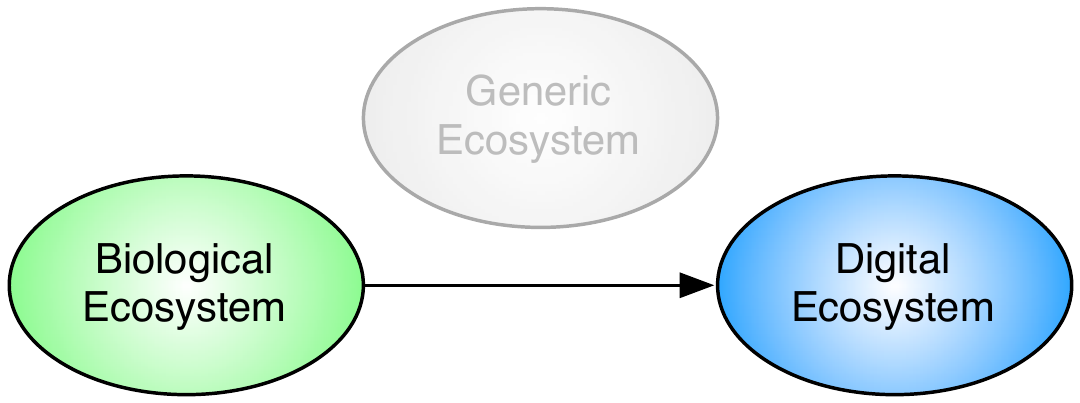}{pdf}{Creation of Digital Ecosystems}{We considered aspects of biological ecosystems, using a direct unidirectional flow of information and models from biological ecosystems to Digital Ecosystems.}{-4mm}{}{}{}

\tfigure{width=3.33in}{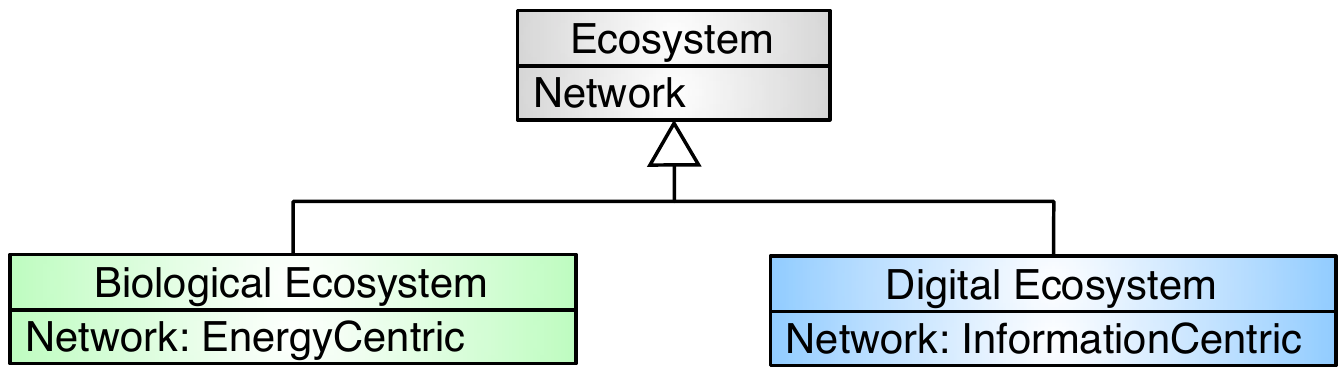}{pdf}{Hypothetical Abstract Ecosystem Definition}{Given an abstract ecosystem class in the \ac{UML}, then the Digital Ecosystem and biological ecosystem classes would both inherit from it, but implement its attributes differently.}{-4mm}{}{}{}

In the creation of Digital Ecosystems \cite{bionetics, de07oz, thesis, acmMedes} we considered aspects of biological ecosystems, including \ac{ABM} \cite{Green2006} and \ac{CAS} \cite{Levin1998}, and then constructed their counterparts in Digital Ecosystems. After which we considered the possibility of a Generic Ecosystem definition \cite{thesis}, because we made use of a direct unidirectional flow of information and models from biological ecosystems to Digital Ecosystems as shown in Figure \ref{figure1}. Without the Generic Ecosystem concept some of the counterparts of biological ecosystems that we constructed in Digital Ecosystems appeared to be compromised, when they were actually the realisation of generic abstract concepts. Most notably the network structure, which is energy-centric in biological ecosystems \cite{Begon1996}, while information-centric in Digital Ecosystems, as shown in Figure \ref{figure2}. So, given an abstract ecosystem class in the \ac{UML}, then the Digital Ecosystem and biological ecosystem classes would both inherit from it, but implement its attributes differently. So, we argued that the apparent compromises in mimicking biological ecosystems were actually features unique to Digital Ecosystems \cite{thesis}. Therefore, there is potential to create a Generic Ecosystem definition, using a suitable modelling technique such as \ac{CAS} \cite{Waldrop1992}, which would abstractly define the key properties of an ecosystem, and would theoretically be applicable to any domain where the modelling technique has been applied. 

We can create a definition of a Generic Ecosystem based on \ac{CAS}, making use of the \ac{ABM} and \acp{MAS},\footnote{An \ac{ABM} is a class of computational models for simulating the actions and interactions of autonomous agents with a view to assessing their effects on the system as a whole. It combines elements of game theory, complex systems, emergence, computational sociology, \acp{MAS}, and evolutionary programming.} which will define the key properties of an ecosystem, in any domain, in an abstract extensible form. Therefore, the Generic Ecosystem definition will provide a framework for the application of ideas, concepts, and models from one class of ecosystem to another, including Digital, Business and Knowledge Ecosystems. Naturally, biological ecosystems will be the main source of information for the conceptualisation of the Generic Ecosystem.

\section{Biological Ecosystems}

In order to create a Generic Ecosystem we will consider biological ecosystems in terms of the key properties, behaviours and structures. These were considered extensively \cite{thesis,de07oz,dbebkpub}, and so we will summarise the main findings.

\tfigure{width=3.33in}{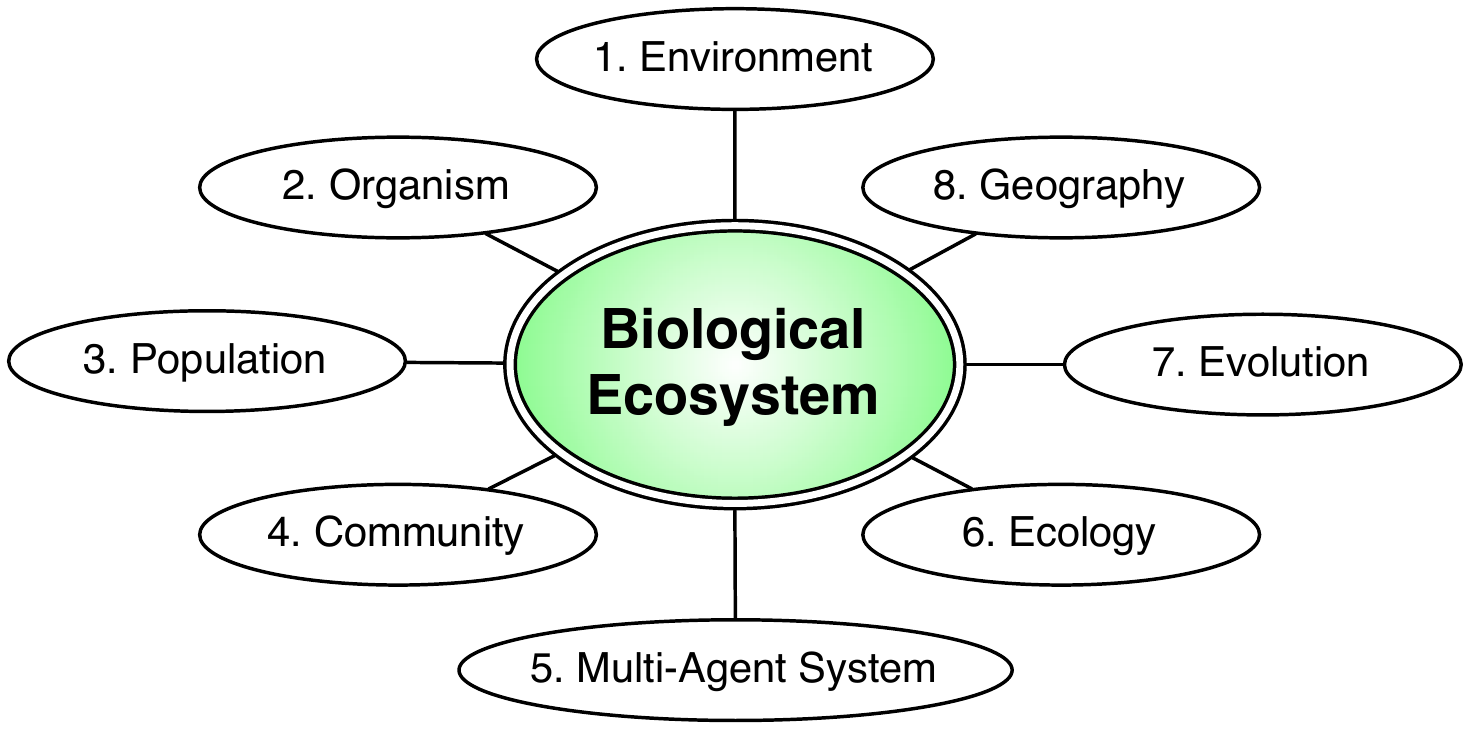}{pdf}{Biological Ecosystem}{Key properties, behaviours and structures based on our understanding from \cite{thesis, de07oz, dbebkpub}. This mind-map will allow for the easy transition of understanding from biological ecosystems to the Generic Ecosystems.}{-4mm}{!b}{}{}

Ecosystems are often described as \ac{CAS}, because like them, they are systems made from diverse, locally interacting components that are subject to selection. Other \ac{CAS} include brains, individuals, economies, and the biosphere. All are characterised by hierarchical organisation, continual adaptation and novelty, and non-equilibrium dynamics. These properties lead to behaviour that is non-linear, historically contingent, subject to thresholds, and contains multiple basins of attraction \cite{Levin1998}. The features of these systems, especially non-linearity and non-equilibrium dynamics, offer both advantages and hazards for adaptive problem-solving. The major hazard is that the dynamics of \ac{CAS} are intrinsically hard to predict because of the non-linear emergent dynamics \cite{Levin1999}. The occurrence of multiple basins of attraction in \ac{CAS} suggests that even a system that functions well for a long period may suddenly at some point transition to a less desirable state \cite{Folke2004}. Non-linear behaviour provides the opportunity for scalable organisation and the evolution of complex hierarchical solutions, while rapid state transitions potentially allow the system to adapt to sudden environmental changes with minimal loss of functionality \cite{Levin1998}. 

In creating Digital Ecosystems, the digital counterpart of biological ecosystems, we naturally asked their likeness to the biological ecosystems from which they came \cite{thesis}. Further to this, we could consider the applicability of other aspects of ecosystems theory in understanding and analysing the dynamics of Digital Ecosystems. For example, energy pyramids\footnote{Energy pyramids show the dissipation of energy at trophic levels, positions that organisms occupy in a food chain, e.g. producers or consumers \cite{Odum1968}.} of biological ecosystems, what is their equivalent in Digital Ecosystems? Given that Digital Ecosystems are information-centric, whereas biological ecosystems are energy-centric \cite{Begon1996}, they would undoubtedly be information pyramids, but further definition would naturally require more research. 

So, we can define a framework for understanding biological ecosystems, with the aim of applying that understanding to Digital Ecosystems, through a Generic Ecosystem definition. Our understanding of a biological ecosystem is summarised as a mind-map in Figure \ref{figure3}. This mind-map will allow for the easy transition of understanding from biological ecosystems to the Generic Ecosystem. It can also be easily extended if and when we find new and relevant understanding, given that research into biological ecosystems is ongoing.

\vspace{8mm}
\section{Generic Ecosystem}

The Generic Ecosystem will provide a framework for the application of ideas, concepts of models from one class of ecosystem to another, which will be fundamental when combining different classes of ecosystems to create and define applied Digital Ecosystems. Biological ecosystems can be considered in terms of \ac{ABM} \cite{Green2006} and \ac{CAS}, leading us to define a Generic Ecosystem in terms of \ac{MAS} and \ac{CAS}, with agents to represent organisms and a network to represent the geographical landscape, as shown in Figure \ref{figure4}. The evolution, change over time, is biological (Darwinian) \cite{Darwin1859}, and not the more general mathematical interpretation often associated with \ac{CAS}. 

\tfigure{width=3.33in}{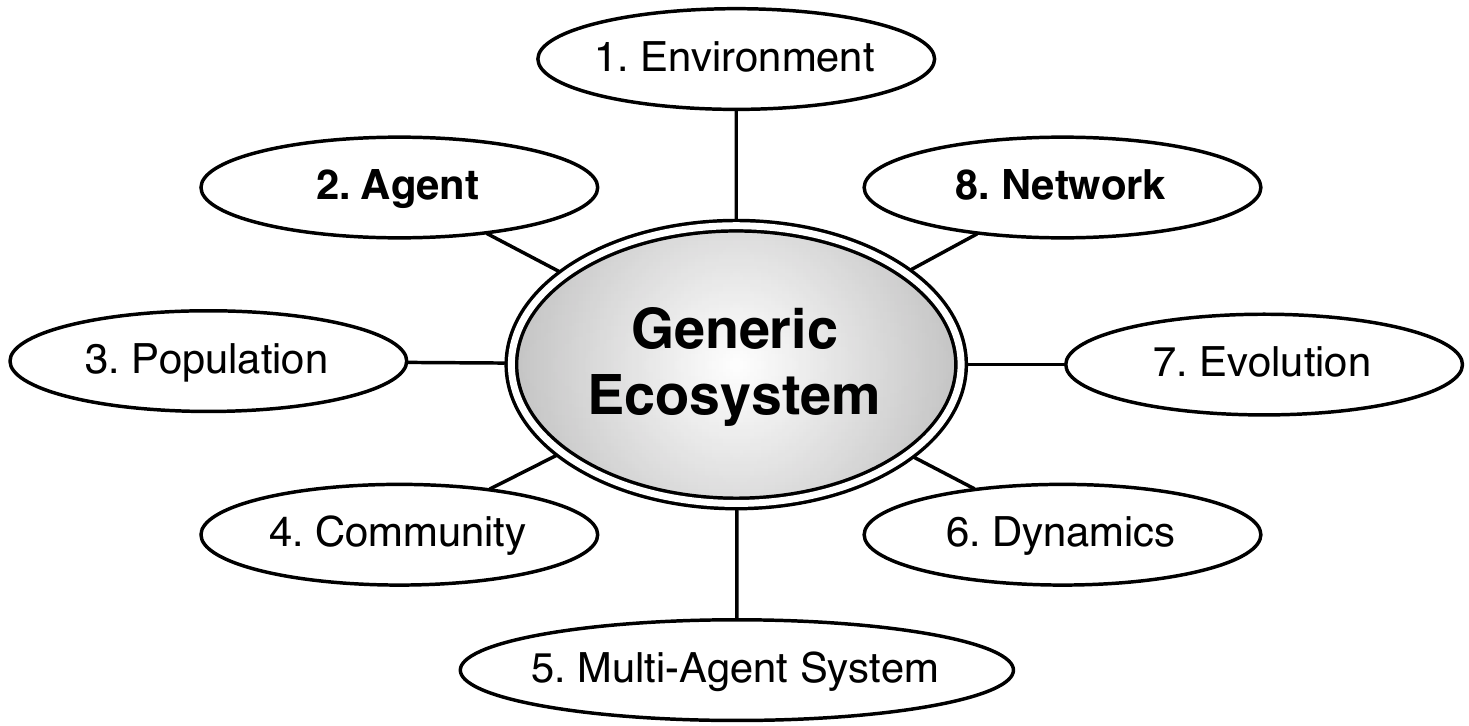}{pdf}{Generic Ecosystem}{Key properties, behaviours and structures based on our understanding from biological ecosystems. The items in bold are the ones that have changed to more generic concepts from biological ecosystems, as defined in Figure \ref{figure3}.}{-4mm}{!b}{}{}

The instantiation of the Generic Ecosystem within a specific domain will create a class of that type of system with ecological properties. While some properties, behaviours, and structures will transition easily between domains, as counterparts already exist or can be easily constructed (e.g. evolution), others will prove more challenging (e.g. ecological dynamics). 

Assuming the motivation for engineering an applied Digital Ecosystem is the development of scalable, adaptive solutions to complex dynamic problems, certain generalisations can be made from biological ecosystems. Sustained diversity \cite{Folke2004}, is a key requirement for dynamic adaptation. In any applied Digital Ecosystem, diversity must be balanced against adaptive efficiency because maintaining large numbers of poorly-adapted solutions is costly. The exact form of this trade off will be guided by the specific requirements of the system in question. Stability \cite{Levin1998}, is likewise, a trade-off: we want the system to respond to environmental change with rapid adaptation, but not to be so responsive that mass extinctions deplete diversity or sudden state changes prevent control. This is an example of the kind of cross ecosystem knowledge transfer to be facilitated, which could be achieved through \acp{BDP}. A design pattern is a general reusable solution to a commonly occurring problem in software design \cite{Gamma1995}. It is not a finished design that can be transformed directly into code, but a description or template for how to solve a problem that can be used in many different situations \cite{Gamma1995}. For example, object-oriented design patterns typically show relationships and interactions between classes or objects, without specifying the final application classes or objects that are involved \cite{Gamma1995}. A \ac{BDP} would extend this concept to catalogue common interactions between biological structures using a pattern-oriented modelling approach \cite{Grimm2005}, which when applied would endow systems with the desirable properties of biological systems, such as self-organisation, self-management, scalability and sustainability.

\subsection{Digital Ecosystem}

\tfigure{width=3.33in}{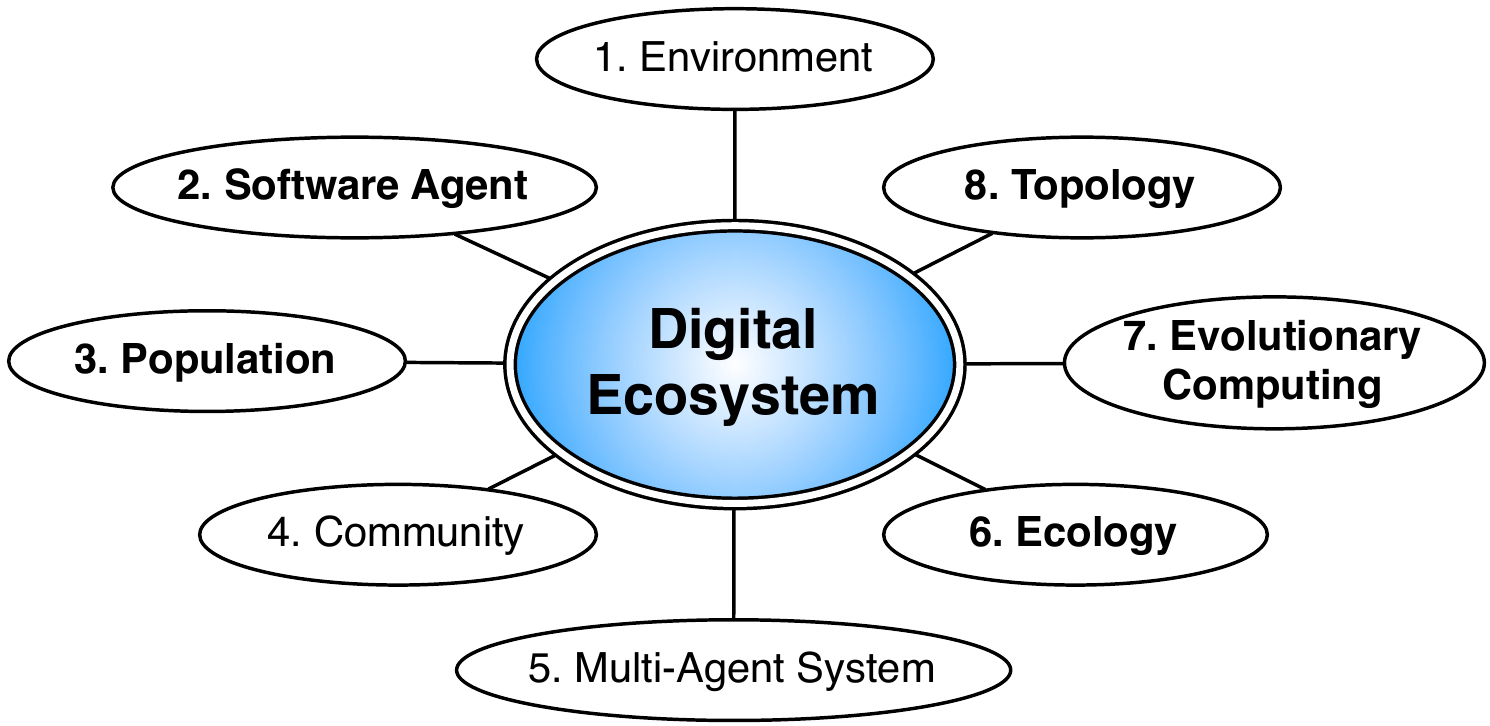}{pdf}{Digital Ecosystem}{Key properties, behaviours and structures based on our understanding of a Generic Ecosystem, with the concepts in bold having changed to more domain specific ones, e.g. \emph{network} to \emph{topology}.}{-4mm}{!b}{}{} 

Figure \ref{figure5} shows the key properties, behaviours and structures of a Digital Ecosystem, based on our understanding of a Generic Ecosystem, with the concepts in bold having changed to more domain specific ones. The concept of \emph{agent} from the Generic Ecosystem naturally maps to \emph{software agents} \cite{bionetics, thesis}. The agents of the Digital Ecosystem are functionally analogous to the organisms of biological ecosystems, including the behaviour of migration and the ability to be evolved \cite{Begon1996}, and will be achieved through using a hybrid of different technologies. The ability to migrate is provided by using the paradigm of agent mobility from mobile agent systems \cite{Pham1998}, with the habitats of the Digital Ecosystem provided by the facilities of agent stations from mobile agent systems \cite{McCabe1994}, i.e. a distributed network of locations to migrate to and from.

The concept of \emph{evolution} easily maps to \emph{evolutionary computing}, and therefore so does the concept of \emph{population}, to which the process of evolution applies \cite{Darwin1859}. However, the specifics of applying evolutionary computing in Digital Ecosystems required some consideration. To evolve high-level software components in Digital Ecosystems, we proposed taking advantage of the native method of software advancement, human developers, and making use of evolutionary computing \cite{Eiben2003} for combinatorial optimisation \cite{Papadimitriou1998} of the available software agents (which represent services). This involves treating developer-produced software services as the functional building blocks, as the base unit in a genetic-algorithms-based process. Furthermore, such an approach requires a modular reusable paradigm to software development, such as \acp{SOA} \cite{Newcomer2005}. 

Mapping the concept of \emph{network} required more effort. Specifically, a distributed information-centric dynamically re-configurable network topology to support the constantly changing multi-objective information-centric \emph{selection pressures} of a user base. This would allow for the connectivity of its habitats to adapt to the connectivity within a user base, with a cluster of habitats representing a community within the user base. So, a network topology that will be discovered with time, and which reflects the connectivity within a user base \cite{thesis, bionetics}.

The mapping of dynamics (ecology) lead to novel form of distributed evolutionary computation for our \ac{EOA} \cite{thesis, bionetics}. This novelty came from the creation of multiple evolving populations in response to \emph{similar} requests, whereas in the island-models of \acl{DEC} there are multiple evolving populations in response to only one request \cite{lin1994cgp}. So, different requests are evaluated on separate \emph{islands} (populations), with their evolution accelerated by the sharing of solutions between the evolving populations (islands), because they are working to solve similar requests (problems).

\subsection{Social Ecosystem}

\tfigure{width=3.33in}{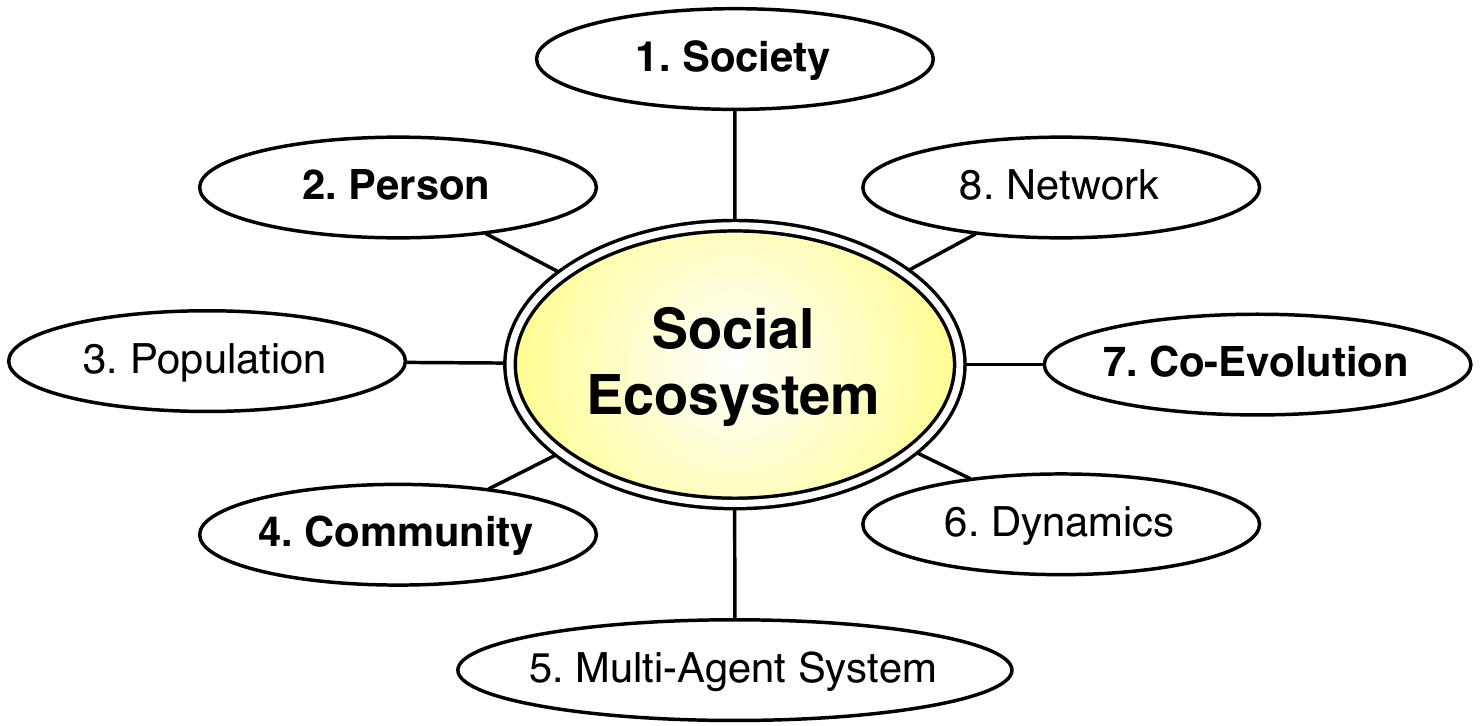}{pdf}{Social Ecosystem}{Key properties, behaviours and structures based on our understanding of a Generic Ecosystem. The concept of \emph{environment} from the Generic Ecosystem maps to that of \emph{society}, while the concept of \emph{agent} maps to that of \emph{person}.}{-4mm}{!b}{}{}

According to Social Ecosystem theory, populations adapt to their environment in order to survive, since it is in the environment where they find the sustenance resources needed for survival, but human populations are the only ones to adapt to their environment through culture \cite{Diez1995}. Therefore, culture may be considered an instrumental response on the part of human populations in order to achieve a better adaptation to their environment \cite{Hawley1986, Diez1983}. Different forms of social organisation constitute cultural responses to the problem of adaptation faced by any population that must survive with the resources which it finds in its environment. So, naturally the concept of \emph{agent} from the Generic Ecosystems maps to that of \emph{person}, and the concept of \emph{environment} maps to that of \emph{society}, as shown in Figure \ref{figure7}.

In biological ecosystems a \emph{community} is a group of interacting populations sharing an environment \cite{Begon1996}. While the concept of \emph{community} is well established within Social Ecosystems, there is no single agreed definition \cite{mcmillan1986sense}. However, in summary, in communities, intent, belief, resources, preferences, needs, risks, and a number of other conditions may be present and common, affecting the identity of the participants and their degree of cohesiveness \cite{mcmillan1986sense}. The word is often used to refer to a group that is organised around common values and is attributed with social cohesion within a shared geographical location, generally in social units larger than a household. The word can also refer to the national community or global community. Since the advent of the Internet, the concept of community no longer has geographical limitations, as people can now virtually gather in an online community and share common interests regardless of physical location. So, we can map the abstract concept of \emph{community} from the Generic Ecosystem to its more domain specific variant.

The concept of \emph{evolution} is mapped to that of \emph{co-evolution}, because the key differentiating point of a Social Ecosystem from a social system is the interdependence among the entities within it, which occurs through the phenomenon of co-evolution \cite{Mitleton2003}. The is consistent biologically, because the environment is society, such that for any person their environment is other people.

In a biological ecosystems, co-evolution is the evolutionary change of an organism triggered by the change of a related organism \cite{Lawrence2005}. Each party in a co-evolutionary relationship exerts selective pressures on the other, thereby affecting each others evolution. Co-evolution may occur in a one-on-one interaction, such as that between predator and prey, host-symbiont or host-parasitic pair, but many cases are less clear-cut; a species may evolve in response to a number of other species, each of which is also evolving in response to a set of species. This situation has been referred to as diffuse co-evolution \cite{Thompson1994}, and for many organisms the biotic (living) environment is the most prominent selective pressure resulting in evolutionary change. We would suggest that the same is true for Social Ecosystems, such that the majority of its co-evolution is also diffuse.

\vspace{1cm}
\subsection{Business Ecosystem}

\tfigure{width=3.33in}{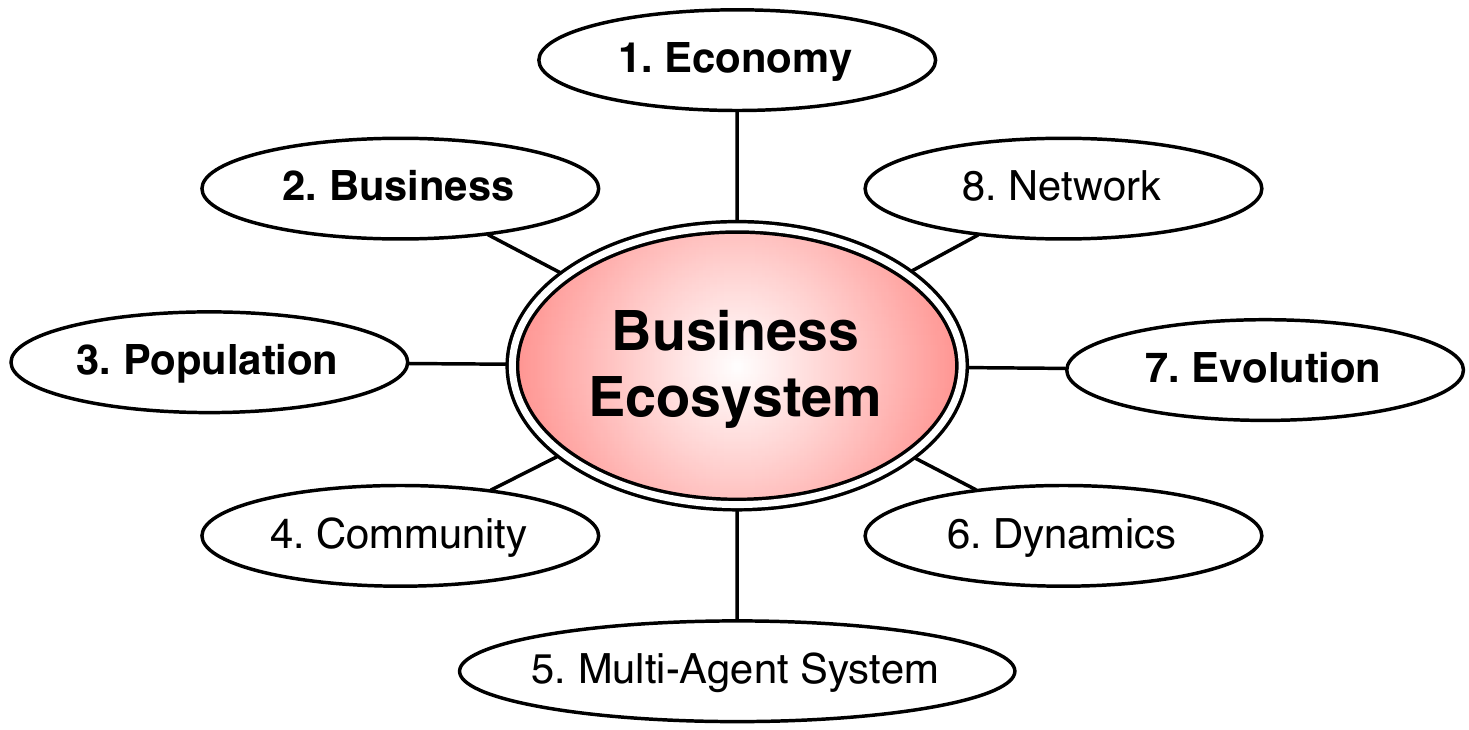}{pdf}{Business Ecosystem}{Key properties, behaviours and structures based on our understanding of a Generic Ecosystem. The concept of \emph{environment} from the Generic Ecosystem maps to that of \emph{economy}.}{-4mm}{}{}{}

The concept of a \emph{business ecosystem} \cite{Moore1996} is well-defined and is focused on the micro-economic view of business networks, whereas the Business Ecosystem has a macro-economic perspective \cite{dbebook}. However, it should not be confused with ecological economics, which is a transdisciplinary field that aims to address the interdependence of human economies and biological ecosystems \cite{Costanza1997}. Therefore, the concept of \emph{environment} from the Generic Ecosystem maps to that of the economy, as shown in Figure \ref{figure6}. While the concept of an \emph{agent} from the Generic Ecosystem naturally maps to that of a \emph{business} in the Business Ecosystem. Each agent (business) is a participant which both influences and is influenced by the environment (economy) of the Business Ecosystem, which is made up of all the businesses, consumers, and suppliers, as well as economic and legal institutions \cite{Mitleton2003}.

Evolutionary theory is well understood within economics \cite{Nelson1982}, so the concept of \emph{evolution} from the Generic Ecosystems can be mapped to its more domain specific variant, as can the concept of \emph{population} to which the process of evolution occurs \cite{Darwin1859}. However, ecosystems theory, including ecological dynamics, is not well understood within economics. We could use our efforts with Digital Ecosystems as a case study, following the same process to define Business Ecosystems. Alternatively, we could instead make use of the Generic Ecosystem definition, because there is extensive work on the \ac{ABM} of economic systems \cite{Tesfatsion2002}, which we can take advantage of in defining a \ac{CAS}/\ac{MAS}-based definition for an Business Ecosystem.

\subsection{Knowledge Ecosystem}

\tfigure{width=3.33in}{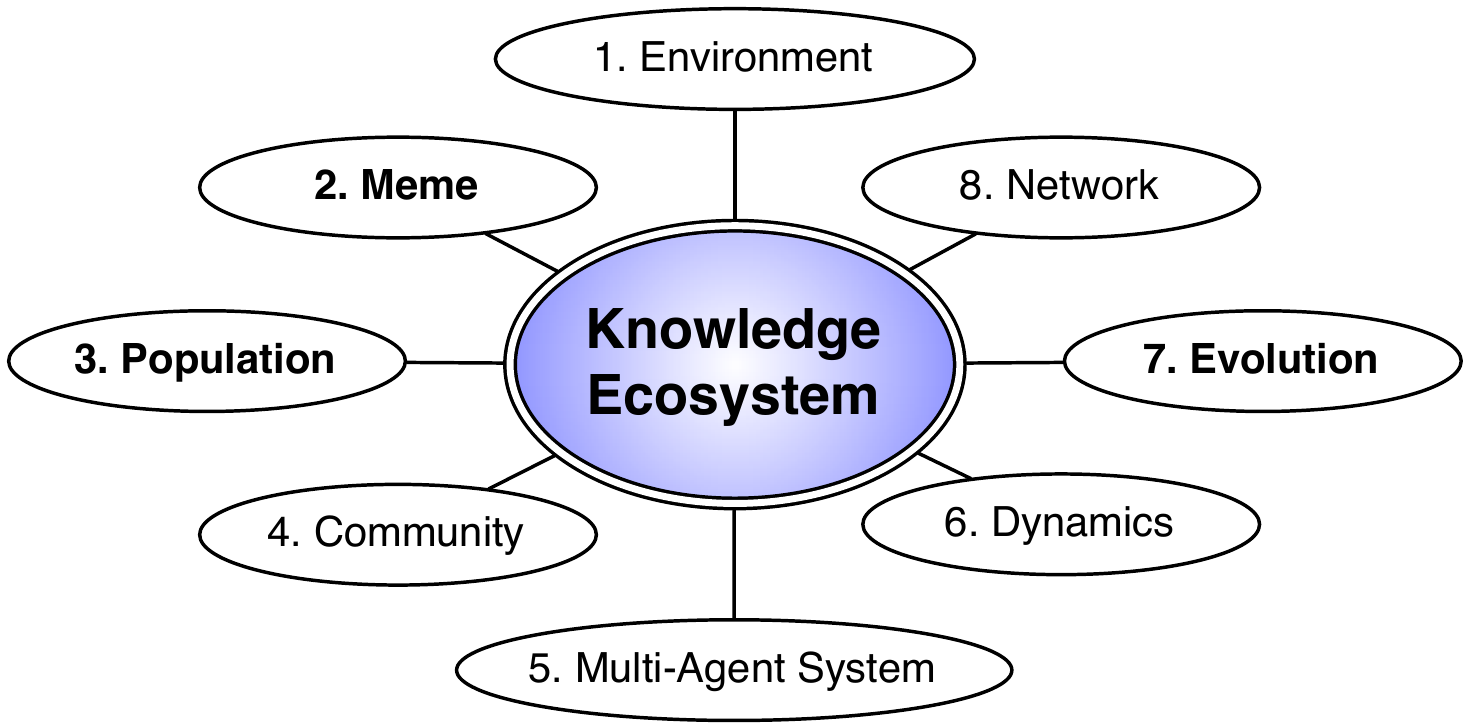}{pdf}{Knowledge Ecosystem}{Key properties, behaviours and structures based on our understanding of a Generic Ecosystem. The concept of \emph{agent} from the Generic Ecosystems maps to that of \emph{meme}.}{-4mm}{!b}{}{} 

An extension of knowledge management ideas, a Knowledge Ecosystem fosters the dynamic evolution of knowledge interactions between entities to improve decision-making and innovation. This bottom-up approach seeks to be more resilient \cite{March1999}. In contrast to directive management efforts that attempt either to manage or direct outcomes, Knowledge Ecosystems espouse that knowledge strategies should focus more on enabling self-organisation in response to changing environments \cite{Clippinger1999}. Articles discussing these ecologically-oriented approaches typically incorporate elements of \ac{CAS} \cite{bowonder2000technology}.

There is no single agreed definition of knowledge, but instead numerous competing theories. Still, one way to consider knowledge constructs is as memes. A meme, as defined within memetic theory, comprises a unit of cultural information, the building block of cultural evolution or diffusion that propagates from one mind to another analogously to the way in which a gene propagates from one organism to another as a unit of genetic information and of biological evolution \cite{Dawkins2006}. So, the concept of \emph{agent} from the Generic Ecosystems would map to that of \emph{meme}. Therefore, with memes, some knowledge will propagate less successfully and become extinct, while others will survive, spread, and, for better or for worse, mutate \cite{Dawkins2006}. Meme theorists contend that memes evolve by natural selection similarly to Darwinian biological evolution through the processes of variation, mutation, competition, and inheritance. So we can map the concept of \emph{evolution} from the Generic Ecosystem to the more domain specific variant, as a well as the concept of \emph{population} to which the process of evolution occurs \cite{Darwin1859}.

\section{Applied Digital Ecosystems}

As Figure \ref{figure10} shows, with this conceptual framework the majority of information flow for defining a Generic Ecosystem comes, unsurprisingly, from biological ecosystems. However, it also allows for the transfer of realised abstract concepts, through the Generic Ecosystem, from one class of ecosystem to another. 

\tfigure{width=3.33in}{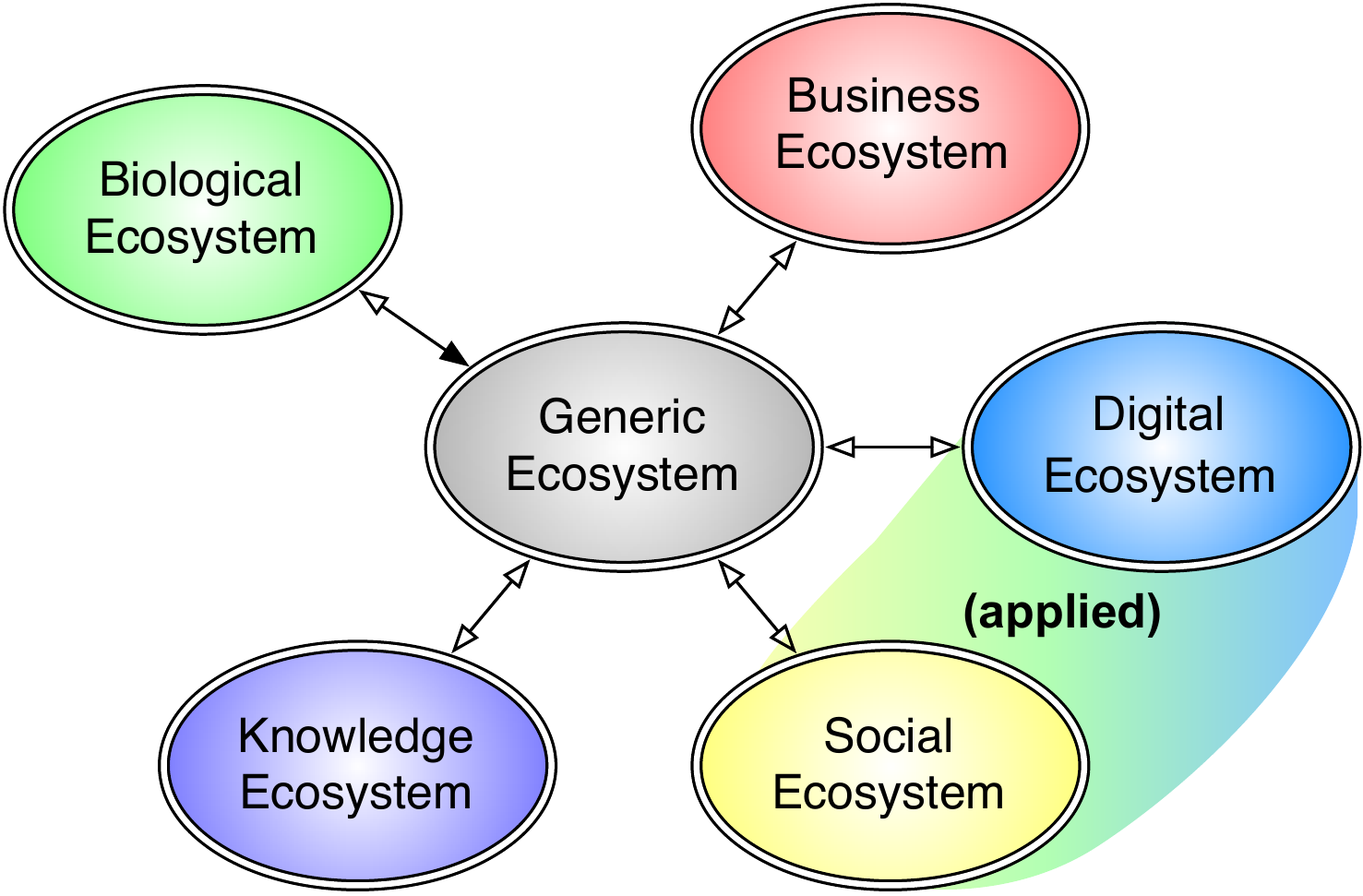}{pdf}{Ecosystems}{The arrows represent information flow between conceptual models of understanding, with the majority coming from biological ecosystems. So, an applied Digital Ecosystem as its combination with a Social Ecosystem.}{-4mm}{!b}{}{}

\tfigure{width=3.33in}{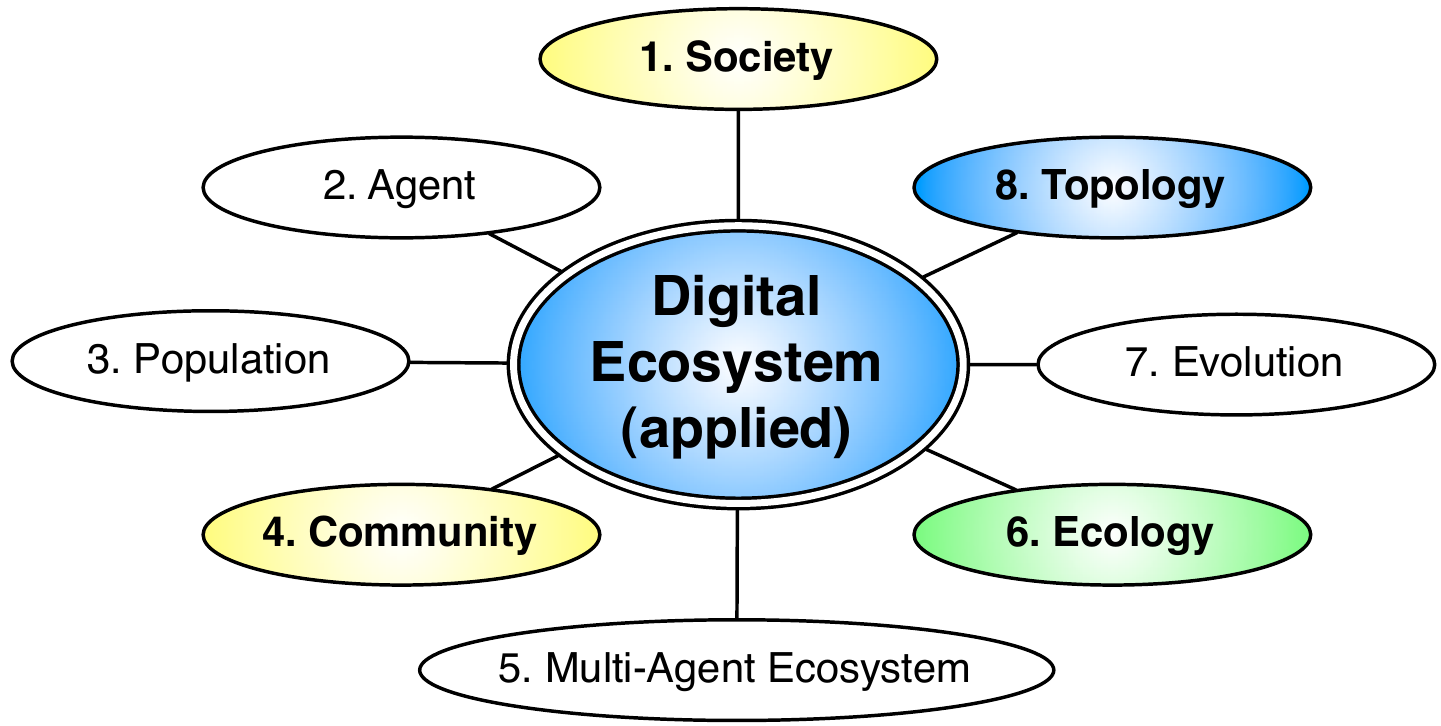}{pdf}{Applied Digital Ecosystem}{Key properties, behaviours and structures based on combining concepts and understanding from Social Ecosystems, Digital Ecosystems, and biological ecosystems through our Generic Ecosystem.}{-4mm}{}{}{}

We can now define an applied Digital Ecosystem as its combination with a Social Ecosystem; therefore, any distributed adaptive open socio-technical system, with properties of self-organisation, scalability and sustainability, inspired by biological ecosystems, as shown in Figure \ref{figure11}. The items in bold are the ones that have changed to more domain specific concepts, with the background colours indicating the class of ecosystem from which the concepts originate. So, an applied Digital Ecosystem adopts the concept of \emph{ecology} from biological ecosystems, the concept of \emph{society} and \emph{community} from Social Ecosystems, and the concept of \emph{topology} from Digital Ecosystems (as we will have a digital information-centric network, rather than a biological energy-centric or a sociological geographically-centric one). The other concepts will depend on the application of the Digital Ecosystem, for example \emph{evolution} from the Generic Ecosystem could map to the \emph{evolution} of biological ecosystems, \emph{evolutionary computing} of Digital Ecosystems, the \emph{co-evolution} of Social Ecosystems, or a domain specific variant of evolution from the application space, or a combination of these. The same therefore also applies to the concept of \emph{agent} and \emph{population}, to which the process of evolution will occur. Also, the concept of \emph{society} will become more specific depending upon the application to which the Digital Ecosystem is applied. This will be further explained as we consider applied Digital Ecosystems which make use of the Business and Knowledge Ecosystems. Furthermore, all these classes of ecosystems can be modelled through \ac{ABM} as \acp{MAS}, allowing us to reasonably combine concepts from these different ecosystems.

\subsection{\acl{DBE}}

The \ac{DBE} is a proposed methodology for economic and technological innovation. Specifically, the \ac{DBE} is a software infrastructure for supporting large numbers of interacting business users and services \cite{dbebkintro}. It aims to be a next generation \ac{ICT} that will extend the \ac{SOA} concept with the automatic combining of available and applicable services in a scalable architecture, to meet business user requests for applications that facilitate business processes. In essence, the \ac{DBE} will be an Internet-based environment in which businesses will be able to interact with each other in very effective and efficient ways \cite{dbebook}. The synthesis of the concept of Digital Business Ecosystems emerged by adding \cite{nachira} \emph{digital} in front of \emph{business ecosystem} \cite{Moore1996}. The term Digital Business Ecosystem was used earlier, but with a focus exclusively on developing countries \cite{Moore2003}. The generalisation of the term to refer to a new interpretation of what \emph{socio-economic development catalysed by \ac{ICT}} means was new, emphasising the co-evolution between the \emph{business ecosystem} and its partial digital representation: the digital ecosystem. The term Digital Business Ecosystem came to represent the combination of these two ecosystems \cite{dbebkintro}.

However, we can now define a \ac{DBE} as a combination of Digital, Social, and Business Ecosystems; therefore, any distributed adaptive open socio-technical system for \emph{business}, with properties of self-organisation, scalability and sustainability, inspired by biological ecosystems. The concept of \emph{environment} from the Generic Ecosystem maps to the \emph{economy} of society. In addition to our definition of applied Digital Ecosystems in the previous section, we have mapped the abstract concept of \emph{agent}, \emph{population} and \emph{evolution} from the Generic Ecosystem to the domain specific concept of \emph{business} and variants of \emph{population} and \emph{evolution}. The mapping proposed not only depends on how the Digital Ecosystem would be implemented, but also how Business Ecosystems are interpreted. For example, some efforts in defining \acp{DBE} are more biased towards to the Digital Ecosystem \cite{bionetics, thesis}, while more recent efforts \cite{abcdbe} are more aligned with this definition.

\tfigure{width=3.33in}{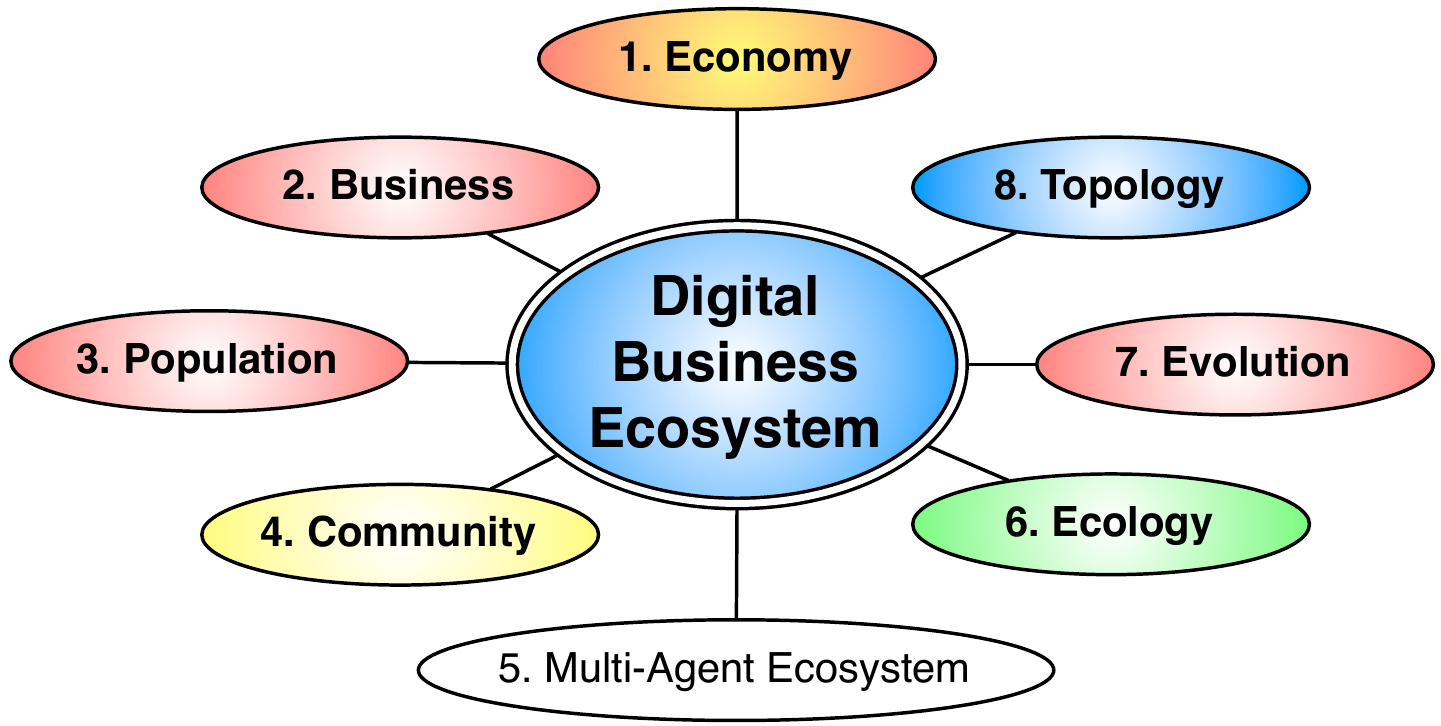}{pdf}{\acl{DBE}}{Key properties, behaviours and structures based on combining concepts and understanding from Social Ecosystems, Digital Ecosystems, and biological ecosystems through our Generic Ecosystem. }{-4mm}{}{1mm}{}

\subsection{\acl{DKE}}

We can now define a \ac{DKE} as a combination of Digital, Social, and Knowledge Ecosystems; therefore, any distributed adaptive open socio-technical system for \emph{knowledge sharing and management}, with properties of self-organisation, scalability and sustainability, inspired by biological ecosystems. The concept of \emph{environment} from the Generic Ecosystem maps to that of society in a \ac{DKE}. In addition to our definition of applied Digital Ecosystems in the previous section, we have mapped the abstract concept of \emph{agent}, \emph{population} and \emph{evolution} from the Generic Ecosystem to the domain specific concept of \emph{meme} and variants of \emph{population} and \emph{evolution}. The mapping proposed not only depends on how the Digital Ecosystem would be implemented, but also how the Knowledge Ecosystem is interpreted. 

Wikipedia and Arxiv.org could be considered \acp{DKE}, because they have many of the necessary properties, except for the \emph{topology} (distributed technical infrastructure). The \ac{DEAL} \cite{DEAL} can also be considered to be a \ac{DKE}, where the knowledge sharing and management is for the benefit of the \emph{society} of rural agriculture. However, the \emph{topology} (distributed technical infrastructure) of Digital Ecosystems is still lacking. In addition to the required distributed technical infrastructure, the necessary legal framework and political support are required for the development and deployment of Digital Ecosystems \cite{abcdbe}. 

\tfigure{width=3.33in}{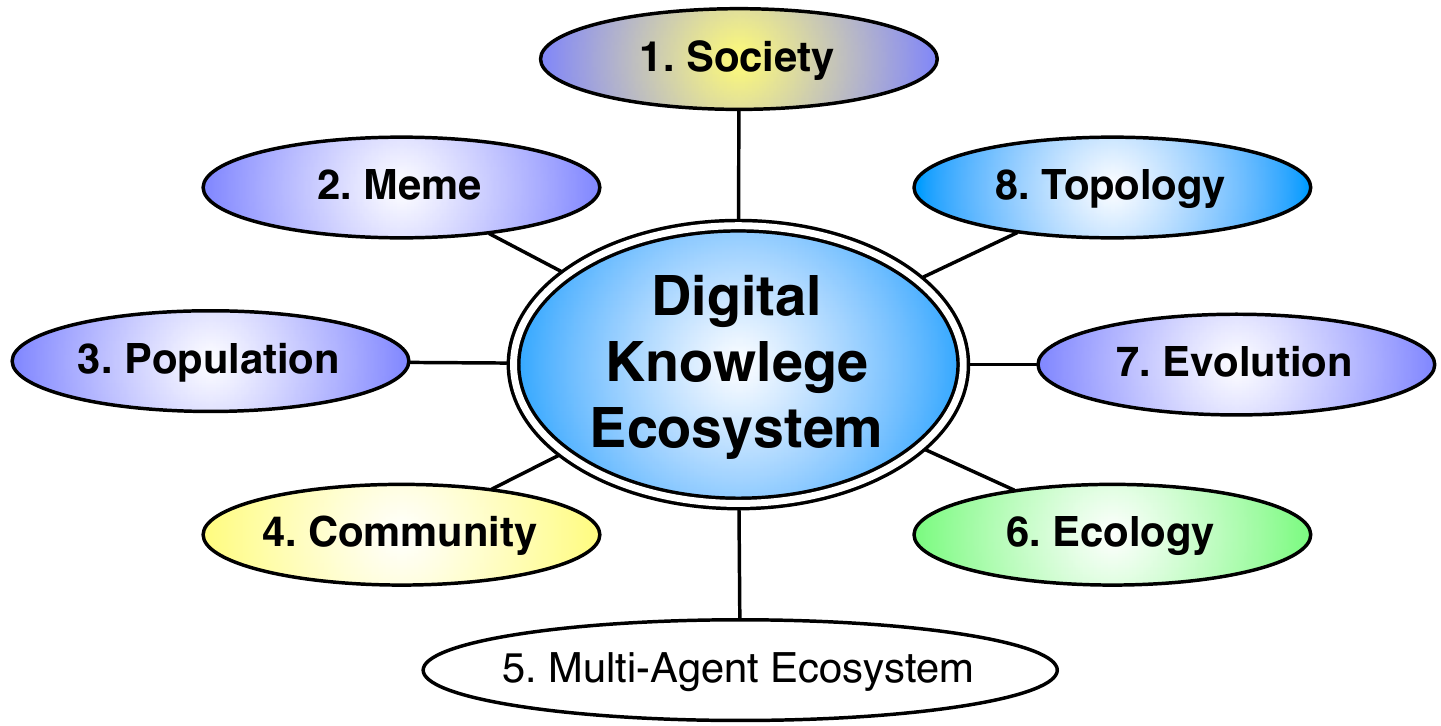}{pdf}{\acl{DKE}}{The key properties, behaviours and structures of a \ac{DKE}, based on combining concepts from Social Ecosystems, Digital Ecosystems, Knowledge Ecosystems and biological ecosystems.}{-4mm}{}{}{}

\section{Conclusions}

We have provided a conceptual and theoretical discussion, from a computer science epistemology, of the characteristics of Digital Ecosystems. Including a discussion of the relevant interfaces between complexity science, sociology, economics and biology to define the nature of Digital Ecosystems and their application. We have provided a conceptual framework for the cross pollination of ideas, concepts and understanding between different classes of ecosystems, based on the universally applicable principles of \ac{ABM} and \ac{CAS}. Using \ac{ABM} to interpret the different classes of ecosystems as \acp{MAS} and therefore facilitate cross-disciplinary understanding between them. Furthermore, we have used this approach to robustly define Digital Ecosystems, including different classes of applied Digital Ecosystems. Therefore, providing a framework to assist cross-disciplinary collaboration in Digital Ecosystems research.

There are of course other dimensions to Digital Ecosystems to be considered, such as the necessary technical infrastructure (i.e. access to the Internet), legal frameworks and political support required for the development and effective deployment of Digital Ecosystems at all levels (economic, social, technical and political). For example, \acp{DBE} to produce real impacts in the economic activities of regions through the improvement of their \ac{SME} business environments \cite{abcdbe}. Also the realisation of Digital Ecosystems in the context of emerging computational paradigms, such as Cloud Computing and Sustainable Computing, which have the potential to radically change the landscape of computational resource provisioning \cite{c3}. For example, Digital Ecosystems risk being subsumed into Cloud Computing at the infrastructure level, while striving for decentralisation at the service level, which would clearly be incompatible with its principles. So, the realisation of the Digital Ecosystems vision requires a form of Cloud Computing, but within the principle of distributed community-based infrastructure, where individual users share ownership \cite{c32}. Sustainable Computing is concerned with achieving environmental sustainability, while abiding by social and ethical responsibilities. So, while Digital Ecosystems would be socially sustainable, there is a lack of a position on environmental sustainability, which is becoming of ever greater importance. We believe that a framework for understanding, such as we propose here, will be required to affectively address these and other, issues and dimensions of Digital Ecosystems.

\section{Acknowledgments}

This work was supported by the EU-funded Open Philosophies for Associative Autopoietic Digital Ecosystems\linebreak (OPAALS) \ac{NoE}, Contract No. FP6/IST-034824.

\bibliographystyle{abbrv}
\bibliography{references,myRefs}
\nocite{briscoeWCAT, boley2007digital} 
\end{document}